\documentclass[aps,prb,twocolumn,amsmath,amssymb,gallery,footinbib,superscriptaddress,floatfix]{revtex4-2}

\usepackage{graphicx}
\usepackage{dcolumn}
\usepackage{bm}
\usepackage{hyperref}
\usepackage{stackrel,amssymb}
\hypersetup{colorlinks=true, linkcolor=blue, citecolor=blue, urlcolor=magenta, pdftitle={Structural chirality measurements and computation of handedness in periodic solids}, pdfauthor={F. Gomez-Ortiz and E. Bousquet}}

\usepackage{fixltx2e}

\usepackage{xcolor} 

\begin{document}

\preprint{APS/123-QED}

\title{Structural chirality measurements and computation of handedness in periodic solids}

\author{Fernando G\'omez-Ortiz}
\email[Corresponding Author:]{fgomez@uliege.be}
\affiliation{Physique Théorique des Matériaux, QMAT, CESAM, Université de Liège, B-4000 Sart-Tilman, Belgium} 
\author{Mauro Fava}
\affiliation{Physique Théorique des Matériaux, QMAT, CESAM, Université de Liège, B-4000 Sart-Tilman, Belgium} 
\author{Emma E. McCabe}
\affiliation{Department of Physics, Durham University, South Road, Durham, DH1 3LE, U. K.}
\author{Aldo H. Romero}
\affiliation{Department of Physics and Astronomy, West Virginia University, Morgantown, WV 26505-6315, USA}
\author{Eric Bousquet}
\email[Corresponding Author:]{eric.bousquet@uliege.be}
\affiliation{Physique Théorique des Matériaux, QMAT, CESAM, Université de Liège, B-4000 Sart-Tilman, Belgium} 
\date{\today}

\begin{abstract}
We compare the various chirality measures most widely used in the literature to quantify chiral symmetry in extended solids, i.e., the continuous chirality measure and the Hausdorff distance.
By studying these functions in an algebraically tractable case, we can evaluate their strengths and weaknesses when applied to more complex crystals. Going beyond those classical calculations, we propose a new method to quantify the handedness of a crystal based on a pseudoscalar function, i.e., the helicity during a soft phonon mode driven displacive phase transition from an achiral structure. 
This quantity, borrowed from hydrodynamics, can be computed from the eigenvector carrying the system from the high-symmetry non-chiral phase to the low-symmetry chiral phase. 
Different model systems like  K$_3$NiO$_2$, CsCuCl$_3$ and MgTi$_2$O$_4$ are used as test cases where we show the superior interest of using helicity to quantify chirality in displacive chiral transitions together with the handedness distinction.
\end{abstract}

\maketitle

\section{Introduction}
\label{sec:introduction}
A recent revival of chirality studies in materials has occurred due to its applicability in high-impact fields like topological insulators or chiral phonons in 2D materials~\cite{Felser-22}. 
Despite this revival, relatively little attention has been paid to the structural chirality in crystalline phases. 
From a symmetry standpoint, the criteria for a crystal to exhibit chirality are straightforward: it must lack improper rotation symmetry elements within its symmetry group (i.e. no "operations of the second kind" which can convert right-handed coordinate systems into left-handed ones)~\cite{Nespolo2018}. 
This condition leads to a comprehensive classification of the 230 crystallographic space groups~\cite{Felser-22,Eric-24}. 
The first category encompasses 165 space groups which include 
improper symmetry operations, rendering them achiral. 
The second category comprises the remaining 65 groups, called Sohncke space groups, which are exclusively characterized by containing orientation-preserving operations (i.e. only "operations of the first kind")~\cite{Nespolo2018}. 
Among the Sohncke groups,  we can further distinguish two different subcategories: 11 enantiomorphic pairs (22 chiral groups) characterized by containing screw axis of opposite handedness and 43 non-enantiomorphic space groups that will only preserve the chirality of the motif. 
The former subset of space groups will be the focus of this work.

However, this binary (chiral vs. achiral) classification sometimes proves insufficient, particularly when we aim to identify materials with optimal chiral responses for specific functionalities such as optoelectronics~\cite{Parker-18}. 
A more nuanced approach involves categorizing crystals based on their varying degrees of chirality, offering a richer classification of materials~\cite{Avnir-92,Avnir-95}.
To address this challenge, significant efforts have been dedicated in the literature to continuously quantify chirality over the past few years despite its inherent difficulty~\cite{Weinberg-00,petijean2003,fowler2005}.
Evidence of such attempts can be found in various measures, including the continuous chirality measure~\cite{Avnir-92,Avnir-95}, the Hausdorff distance~\cite{Buda-92,Felser-22}, or the phonon pseudo-angular momentum of the eigendisplacements~\cite{Streib-21,Zhang-14}.
Chirality measures are commonly divided into two primary categories based on their mathematical behavior under symmetry transformations: scalar and pseudoscalar functions. Scalar functions retain their value when subjected to mirror symmetry, making them invariant under such transformations. In contrast, pseudoscalar functions are sensitive to these symmetries and undergo a sign change when reflected. Intriguingly, in structural chirality determination, scalar functions are predominantly utilized in most established measures. This prevalence overshadows pseudoscalar functions, which remain relatively unexplored and elusive to the best of our knowledge.

In this study, we delve into the strengths and weaknesses of the previously announced chirality measures, e.g., the continuous chirality measure and the Hausdorff distance.
Going further, we propose a fresh methodology to determine the handedness of crystalline solids that undergo displacive chiral transitions \cite{Fava-24} by utilizing a pseudoscalar function inspired by the concept of helicity in hydrodynamics~\cite{Moffat-92,Moffatt2014}. 
This innovative approach quantifies the degree of chirality in a crystal and distinguishes between right- and left-handed configurations. 

However, it is worth noting that while all handed objects are inherently chiral, not all chiral objects possess unambiguous handedness, even if we can distinguish their enantiomers ~\cite{Felser-22}. 
Consequently, our proposed method effectively assesses the chiral response of crystals undergoing a displacive chiral transition within the 11 pairs of enantiomorphic space groups~\cite{Nespolo-21}, albeit it poses problems to those within chiral non-enantiomorphic space groups, limiting its applicability.

The article is organized as follows. 
In the following section, we will inquire into the highly tractable scenario of a unit perovskite ABO$_3$ cell, where all the various chiral measurements can be computed algebraically, providing a clear understanding of their limitations. 
Moreover, we shall introduce the concept of helicity as a new chiral measure to quantify the handedness of a crystalline structure that undergoes a displacive chiral transition. 
After identifying the strengths and weaknesses of the aforementioned quantities, we shall apply them to different cases of interest. 
Subsequent sections will therefore be dedicated to the particular study of K$_3$NiO$_2$, Na$_3$AuO$_2$, CsCuCl$_3$, and MgTi$_2$O$_4$ that have been selected as paradigmatic examples.
\section{The case of a unit perovskite cell}
In this section, we present the continuous chirality measure and the Hausdorff distance through a simple toy model formed by an isolated ABO$_3$ perovskite unit cell.
This examination aims to delineate the applicability range of these approaches. 
Additionally, we shall introduce a new pseudoscalar measure to quantify the handedness of a given structure precisely.
\subsection{Problems with continuous chirality measurements}
\label{Sec:problems}

\begin{center}
  \begin{figure}[h]
     \centering
      \includegraphics[width=\columnwidth]{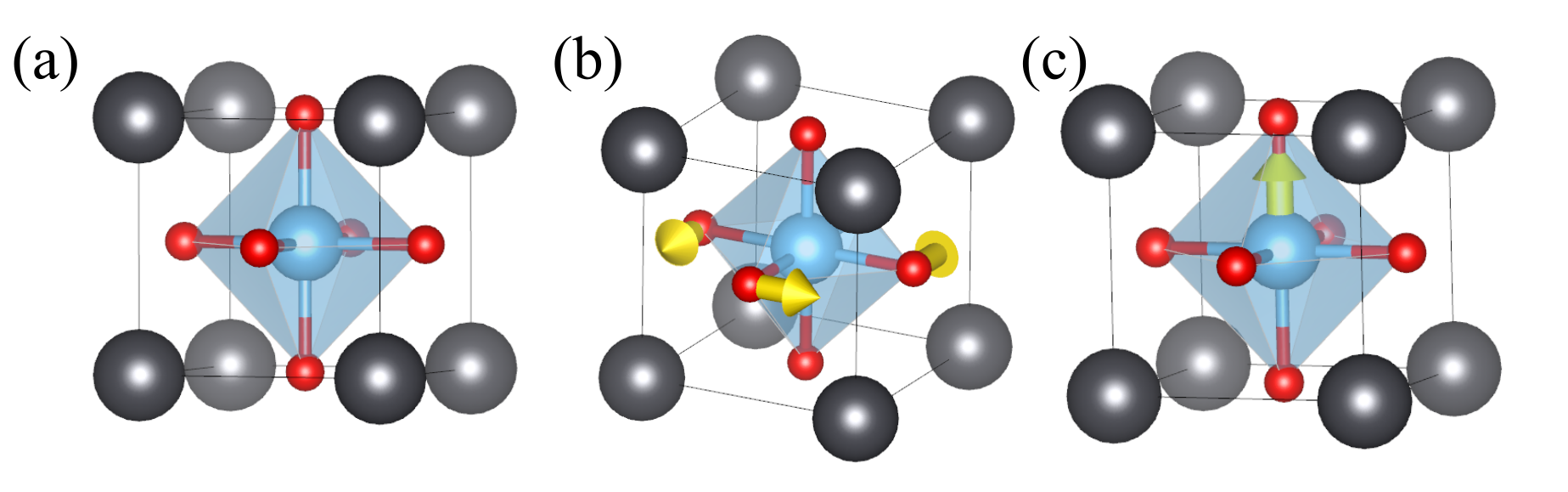}
      \caption{Schematic view of the unit cell in the prototypical cubic ABO$_3$ perovskite structure together with their dominant instabilities without periodic boundary conditions. (a) Non-polar cubic centrosymmetric phase. (b) Non-polar rotation of the oxygen octahedra. (c) The off-centering motion of the B atom. Grey, blue, and red balls represent A, B, and O atoms, respectively. Yellow arrows indicate the direction for the non-zero displacements.}
      \label{fig:distortions} 
  \end{figure}
\end{center}

We begin by examining a simplified model system: a prototypical ABO$_3$ perovskite unit cell, without considering its periodic replicas. In its centrosymmetric cubic phase, corresponding to the $Pm\bar{3}m$ space group under hypothetical periodic boundary conditions, the crystal is inherently non-chiral, as illustrated in Fig.~\ref{fig:distortions}(a). Next, we introduce structural distortions to this system and analyze their impact on different chirality measurements.
According to Hlinka, chirality can arise from the interaction of a vector $V$ with a rotational mode $R$, forming an axial vector. Although these distortions, when considered separately, might be achiral, their coupling can lead to the emergence of chiral distortions~\cite{Hlinka-14}.

These vector and rotation modes can be achieved in our model by the distortions schematized in Fig.~\ref{fig:distortions}.  
In particular, a local rotation in a perovskite can be obtained by a rotation of the oxygen octahedra [Fig.~\ref{fig:distortions}(b)], and a polar vector can be simplified by an off-centering of the B cation [Fig.~\ref{fig:distortions}(c)]. 
Such transformations are easy to characterize algebraically, and the values of the new positions of the distorted atoms can be found in the following Eq.~(\ref{eq:distortion}):
\begin{align}
    O_{\rm x1} &= (\frac{1}{2}-\frac{1}{2}\cos\theta,\frac{1}{2}-\frac{1}{2}\sin\theta,\frac{1}{2}) ,\nonumber\\
    O_{\rm x2} &= (\frac{1}{2}+\frac{1}{2}\cos\theta,\frac{1}{2}+\frac{1}{2}\sin\theta,\frac{1}{2}) ,\nonumber\\
    O_{\rm y1} &= (\frac{1}{2}+\frac{1}{2}\sin\theta,\frac{1}{2}-\frac{1}{2}\cos\theta,\frac{1}{2}) ,\nonumber\\
    O_{\rm y2} &= (\frac{1}{2}-\frac{1}{2}\sin\theta,\frac{1}{2}+\frac{1}{2}\cos\theta,\frac{1}{2}) ,\nonumber\\
    B &= (\frac{1}{2},\frac{1}{2},\frac{1}{2}+\xi),
    \label{eq:distortion}
\end{align}
where $\theta$ is the rotation angle of the oxygen octahedra with respect to the $z$-axis and $\xi$ is the off-centering of the B atom. 
We have, therefore, three different structures: 
i) the high-symmetry cubic phase (that would correspond to a $Pm\bar{3}m$ space group with periodic boundary conditions), 
ii) the lower symmetry phase corresponding to the rotation of the oxygen octahedra (represented by a $I4/mcm$ space group with periodic boundary conditions), and 
iii) the phase corresponding to the off-centering of the B cation (ascribed to a $I4/mcm$ space group with periodic boundary conditions). 
Clearly, the two distortions are nonchiral individually, as they present mirror symmetry planes along $z$ and $x$, respectively. 
However, when the octahedral rotation is combined with the off-centering motion of the B cation, a chiral structure emerges at the unit cell level as it combines a vector (polar distortion) and an axial vector (octahedra rotation). 
It is important to highlight that, in the periodic crystal, the antiferrodistortive coupling of the oxygen octahedra causes octahedra to rotate cooperatively in opposite directions from unit cell to unit cell. In contrast, the off-centering direction is preserved, leading to opposite chirality and an overall achiral system. 
The system would be antichiral unit-cell wise as the chirality from one unit cell will reverse sign but keep the same absolute value concerning its neighboring cells.
Nevertheless, for the sake of simplicity in our analysis, we will restrict our examination to the chiral nature observed at the unit cell level in this straightforward scenario.
\subsubsection{Continuous Chirality Measure}
\label{Sec:prob_CCM}
The continuous chirality measure (CCM) is a scalar measure introduced by Avnir that quantifies the structure's distance to its closest achiral reference~\cite{Avnir-95}. 
Mathematically, it can be expressed as:
\begin{equation}
    {\rm{CCM}}=\frac{1}{N}\sum\limits_1^N \vert\vert\vec{x}_i-\vec{x}_i^\prime\vert\vert^2,
    \label{eq:ccm}
\end{equation}
where $N$ is the number of atoms of the structure and $\vec{x},~\vec{x}^\prime$ are respectively the positions of the structure under study and its closest achiral reference. Although in later works, Avnir added a normalization factor by the root mean square size of the original centered structure~\cite{Dryzun-11} this will not be considered here for simplicity.

One inherent difficulty of the CCM~\cite{Avnir-92,Avnir-95} is that it requires selecting the closest non-chiral structure. 
One may think that a reliable choice would be the non-distorted high symmetry phase, arguing that for small enough distortions, the closest non-chiral configuration is the original reference structure. 
This simple toy model demonstrates that such an assumption is generally incorrect. 
In the general case, we can follow a symmetry adapted modes (SAMs) argument. Given a nonchiral high-symmetry phase (R) and a chiral low-symmetry phase (G), one can generally write the total distortion $\delta$ as the sum over SAMs, namely as a$\delta_1$ + b$\delta_2$ where $\delta_1$ and $\delta_2$ are the modes associated with the irreducible representations brought up by the R $\rightarrow$ G transition and a and b are real projection coefficients. We can take $\delta_2$ to be the symmetry breaking mode, while $\delta_1$ is the mode associated with an isotropy representation of the parent achiral group. Thus the CCM is given by the norm of $\delta_1$ rather than by the norm of $\delta$ that would be grater.
In fact, the closest non-chiral structure in this example corresponds to the configuration with the octahedral rotation but without the B atom's off-center motion. 
This problem manifests when computing the CCM with the cubic centrosymmetric structure as a reference. 
Applying Eq.~(\ref{eq:ccm}) to our perovskite toy model system with such a reference results in the following value:
\begin{equation}
    {\rm{CCM}}=\frac{1}{15}\sum\limits_{i\in~\rm{atoms}} \vert\vert \vec{x}_i-\vec{x}_{i}^{Pm\bar{3}m}\vert\vert^2=\frac{1}{15}(\xi^2+4\sin^2\frac{\theta}{2}).
\end{equation}
We can see that the CCM decouples the off-center motion of the B atom and the octahedra rotation. 
Therefore, we would obtain a non-zero value if $\theta$ is non-zero and $\xi$ is zero or vice-versa, even if we know that we need both distortions simultaneously to have a chiral structure. 
This is a consequence of the wrong selection of the reference structure. 
However, if we take as a reference the configuration containing the octahedral rotations, which is the closest non-chiral structure, the CCM would then have a value of
\begin{equation}
    {\rm{CCM}}=\frac{1}{15}\sum\limits_{i\in~\rm{atoms}} \vert\vert \vec{x}_i-\vec{x}_{i}^{I4/mcm}\vert\vert^2=\frac{1}{15}\xi^2,
\end{equation}
which is not satisfactory either, as much important information is encoded in the selection of the reference structure, and the value of the final order parameter only relies on the off-centering value of the B atom. 
Consequently, the chirality contribution measure of two different distortions, $D_1$ and $D_2$, will be higher for the distortion with greater off-centering, regardless of the extent of their octahedral rotations. This discrepancy arises because the reference structures for both distortions differ, potentially leading to misconceptions.

These challenges become more pronounced as the crystal structure becomes more complicated. Identifying the closest non-chiral reference structure can be difficult. While recent numerical approximations, available at~\cite{Zayit-web,Tuvi-Arad-24}, have improved computational efficiency to $N^2$ (compared with the $N!$ scaling of analytical methods~\cite{Pinsky-08})  with an error margin of $2\%$ for cyclic symmetry groups~\cite{Dryzun-11}, they may disrupt the connectivity map of the structure, resulting in unphysical references~\cite{Alon-23}. Furthermore, chiral distortions often involve multiple coupled modes in complex systems with many atoms in the unit cell. These modes can become decoupled, as seen in the case where octahedral rotation and the off-centering motion of the B cation are examined separately.

Besides these problem-specific issues, other general problems and considerations about the difficulties of whether continuous chiral measures can be well-defined have been addressed in Ref.~\cite{buda1992,gillat1994,petijean2003,fowler2005,jenkins2018}.
\subsubsection{Hausdorff Distances}
The Hausdorff distance is defined as the supremum of the minimum distances between the structure and its closest non-chiral reference~\cite{Buda-92,Felser-22}. 
Therefore, it also relies on the pre-assumption of a reference structure concerning which computes the supremum of the infimum of the distances (i.e. the supremum for the case of maximal overlap between the structure and the reference). 
Consequently, the same types of problems discussed above for the CCM are at play too for the Hausdorff distance. 
When we compute the Hausdorff distance concerning the centrosymmetric structure, we obtain the following value
\begin{equation}
    \mathbb{H}(chir,Pm\bar{3}m)=\sup \lbrace\vert\xi\vert,\sin\frac{\vert\theta\vert}{2}\rbrace,
\end{equation}
depending on the values of the off-centering motion and the octahedral rotation, the Hausdorff distance would vary. 
However, similar to the case of the CCM, since we are computing positive definite distances, no distinction can be made between enantiomers if we reverse either the direction of the rotation or the direction of the off-centering motion. 
This problematic is common to all scalar measures as mirror symmetries will leave the value unaffected. 

Another problem arises With the Hausdorff distance when considering the distances' supremum. 
For small distortions, the distance between the B atoms is $\xi$, and the distance between the equatorial oxygens is $\sin\frac{\theta}{2}\sim \frac{\theta}{2}$. 
Let us imagine that the value of the off-centering and the rotation is the same. 
In such a case, the Hausdorff distance would be $\xi$. However, the multiplicity of the oxygen atoms is not taken into account. 
Therefore, even if the overall contribution of the oxygen sites is larger than the one coming from the B atom, as the individual inputs are smaller, they are not considered in the chiral measure. 
This is an added problem to the ones highlighted above for the case of the CCM.
\subsection{Computation of Handedness}
\label{sec:comp_Handedness}
Handedness is intrinsically linked to rotational direction in physics. For instance, in chemistry, helicity refers to the sense of rotation of helical structures, with right-handed helices assigned a positive helicity value and left-handed helices assigned a negative helicity value~\cite{Moss-96}. Similarly, in hydrodynamics, the handedness of a flow is determined by the sign of its helicity, which measures the degree of linkage of the streamlines~\cite{Moffatt2014}.
This quantity can be computed from the velocity lines of the flow as the following integral~\cite{Moffat-92,Moffatt2014}:
\begin{equation}
    \mathcal{H}=\int d^3\vec{r}~\vec{v}\cdot\left[\vec{\nabla}\times\vec{v}\right].
    \label{eq:Helicity}
\end{equation}
The integrand of Eq.~(\ref{eq:Helicity}), the helicity density, is a pseudoscalar quantity that changes its sign under a mirror symmetry operation. 
Thus, a nonzero helicity is associated with a lack of mirror symmetry: right (respectively left) handedness can be associated with positive (respectively negative) values of $\mathcal{H}$. 
Accordingly, the helicity modulus, $\vert\mathcal{H}\vert$, quantifies the strength of the handedness~\cite{Moffatt2014}. 
Contrary to scalar measures (like CCM or Hausdorff distances), the helicity measure can discriminate between enantiomers.
However, other problems arise due to the chiral connectedness property~\cite{Mezey-95,Weinberg-97,Banik-16,Vavilin-22}. 
This property refers to the fact that two enantiomorphs can be transformed into one another while remaining chiral throughout the transformation process.
As elucidated in Ref.~\cite{Mezey-95,Weinberg-97,Banik-16}, this inevitably leads to what is known as the \emph{false zeros problem}: Any pseudoscalar function will, in general, fail to determine the handedness of some objects assigning zero values to chiral entities.
A notable exception to the false-zeros problem occurs in the set of helices with variable pitch, as discussed in Ref.~\cite{Mislow-99}. In these cases, pseudoscalar functions exclusively yield zero values for achiral objects, making them suitable measures of chirality. Consequently, enantiomorphic space groups are unaffected by this issue and can be classified according to their handedness, indicating that pseudoscalar functions are more convenient for such classifications.
Conversely, non-enantiomorphic space groups showing chirality like the Pb$_5$Ge$_3$O$_{11}$ compound~\cite{Fava-23}, may exhibit chiral connectedness, making the presence of chiral non-handed configurations inevitable. 

The definition of Helicity from fluid dynamics has been extended to the case of discrete fields defined in periodic crystals to compute the handedness of topological polar textures and is now widely accepted~\cite{Shafer-18,Junquera-23}. Borrowing this knowledge, we shall try to apply it to our particular problem where a velocity field can be established linking the achiral and chiral phases i.e., going from continuous to discrete vector fields. In such a case, the velocity field can be defined, by the individual atomic displacements from the high-symmetry phase to the low-symmetry chiral phase (the difference in positions can be related with a velocity in an arbitrary units of time system following a finite difference approach).
Indeed, we will choose the high-symmetry reference as the non-excited phase and follow the eigenvector direction towards the non-excited chiral state in an arbitrary time length.
Therefore, in our perovskite toy model system, the velocities of the atoms gives the following values:
\begin{align}
    \dot{\vec{x}}_{O_{\rm x1}} &= (\frac{1}{2}-\frac{1}{2}\cos\theta,-\frac{1}{2}\sin\theta,0) ,\nonumber\\
    \dot{\vec{x}}_{O_{\rm x2}} &= (-\frac{1}{2}+\frac{1}{2}\cos\theta,\frac{1}{2}\sin\theta,0) ,\nonumber\\
    \dot{\vec{x}}_{O_{\rm y1}} &= (\frac{1}{2}\sin\theta,\frac{1}{2}-\frac{1}{2}\cos\theta,0) ,\nonumber\\
    \dot{\vec{x}}_{O_{\rm y2}} &= (-\frac{1}{2}\sin\theta,-\frac{1}{2}+\frac{1}{2}\cos\theta,0) ,\nonumber\\
    \dot{\vec{x}}_B &= (0,0,\xi).
    \label{eq:velocities}
\end{align}

The curl $\vec{\nabla}\times\dot{\vec{x}}_i$, is then computed following a finite difference approach and the integral is substituted by a regular sum at each atomic site in the unit cell analogous to the approach employed in polar textures at the supercell level~\cite{Shafer-18,Junquera-23} where the polarization vector field is also discrete. Note that whenever this field cannot be established as it is the case for instance of reconstructive phase transition this method cannot be applied. Furthermore, in the case of a sequence of displacive phase transitions A$\rightarrow$B$\rightarrow$C, where both A and B are achiral and C is chiral, a stepwise approach should be adopted. This approach relies on well-defined eigenvectors governing each individual transition. Consequently, the helicity between the achiral states A$\rightarrow$B would be zero, and the helicity would only be nonzero for the transition B$\rightarrow$C, where chirality emerges.
With this definition and the set of velocities described in Eq.~(\ref{eq:velocities}), we obtain a value of the helicity of:
\begin{equation}
\mathcal{H}=\sin\theta\cdot\xi. 
\end{equation}
Therefore, to have a non-zero helicity, we need the coupling of the oxygen rotation and the polar displacement, as desired, to characterize a chiral structure. 
Moreover, suppose we reverse the sense of rotation or the sign of the polar distortion inducing a change of handedness in the structure. In that case, the helicity changes sign accordingly in contrast to the scenario observed with the CCM or Hausdorff distances. 

In the following sections, we shall apply this method (available at~\cite{github}) to compute the handedness of different crystalline structures that undergo such continuous chiral transitions.
\section{Application to K\textsubscript{3}N\MakeLowercase{i}O\textsubscript{2} like chiral structure}
\label{sec:Na3AuO2}
\begin{center}
  \begin{figure}[h]
     \centering
      \includegraphics[width=6cm]{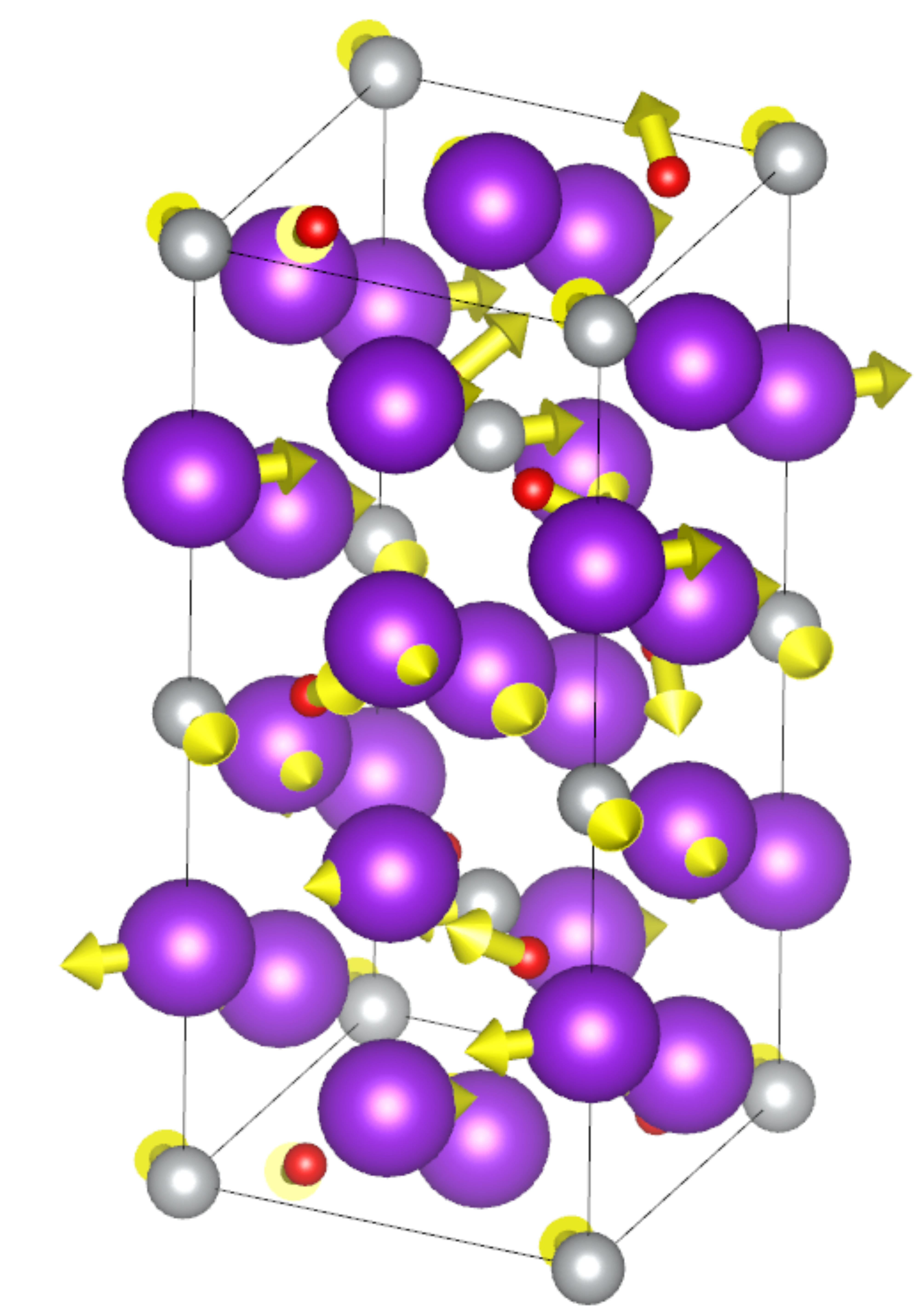}
      \caption{Schematic representation of the K$_3$NiO$_2$ crystal structure where atoms occupy the $P4_2/mnm$ high symmetry positions. 
      Arrows indicate the direction of the atom displacements that bring the system to the $P4_32_12$ chiral phase. 
      Purple, grey, and red balls represent K, Ni, and O atoms} 
      \label{fig:dist_Na3AuO2} 
  \end{figure}
\end{center}

One interesting application case is found in the K$_3$NiO$_2$ compound~\cite{Duris-12}. 
At 423 K, this crystal exhibits a first-order chiral phase transition from a high-symmetry achiral phase ($P4_2/mnm$ space group) to a low-symmetry enantiomorphic phase (either $P4_12_12$ or $P4_32_12$). 
The transition is continuous and has been identified to come from a zone boundary soft phonon mode that explains the cell doubling during the transition~\cite{Fava-24}.
Such a phenomenon could be present in all crystals that can be stabilized into the same high-symmetry $P4_2/mnm$ crystal structure.
For example, although Na$_3$AuO$_2$ does not crystallize in this structure, a hypothetical Na$_3$AuO$_2$ phase isostructural to K$_3$NiO$_2$ is a useful non-magnetic and closed shell toy model system for computational studies.
In this section, we shall compute the CCM, Hausdorff distances and helicity for K$_3$NiO$_2$ and Na$_3$AuO$_2$.
The latter is much simpler from a computational perspective due to its non-magnetic nature. It will be used to systematically analyze how the different chirality measures behave when connecting the different possible phases.
In Fig.~\ref{fig:dist_Na3AuO2}, we show a schematic representation of the K$_3$NiO$_2$ crystal. 
Atomic positions correspond to the high-symmetry $P4_2/mnm$ phase, and the arrows indicate how the atoms move when transitioning to the chiral $P4_12_12$ phase. 
In that way, we have defined a vector field for every atomic site in the reference structure, and therefore, we can compute all the quantities discussed in previous sections. 
We can see from the arrows plotted in Fig.~\ref{fig:dist_Na3AuO2} that a handed helical distortion pattern along the $c$-direction is present.

\begin{table}[htb!]
\caption{\label{tab:dataNa3AuO2} Values of the different chirality measures discussed before for the Na$_3$AuO$_2$ compound taking the $P4_2/mnm$ phase as a reference. The $+$ and $-$ superscripts represent the enantiomorphic $P4_12_12$ and $P4_32_12$ phases, respectively. 
Fractional units and normalization by the number of atoms in the unit cell have been used to compute the CCM and helicity.}
\begin{tabular}{c  c  c}
\hline
   & Na$_3$AuO$_2^+$ & Na$_3$AuO$_2^-$ \\ \hline
CCM & 3.30e-3 & 3.30e-3\\
Hausdorff  &1.18e-1 &1.18e-1\\
Helicity &7.75e-3 &-7.75e-3\\ 
\hline
\end{tabular}
\end{table}

As we can see in Tab.~\ref{tab:dataNa3AuO2} for the case of Na$_3$AuO$_2$, the values of the CCM and Hausdorff distances calculations give each the same value for both enantiomers, which is in line with what we discussed before in Sec.~\ref{Sec:problems}, i.e. that these measures cannot distinguish two enantiomers as distances are always positive definite. 

Finally, the sign of helicity is sensitive to the system's change of handedness while its modulus is constant. 
Hence, and as anticipated in Sec.~\ref{sec:comp_Handedness}, the helicity modulus can measure the chiral distortion amplitude while its sign encodes the handedness ``sign'' of the structure.

\begin{center}
  \begin{figure}[h]
     \centering
      \includegraphics[width=6cm]{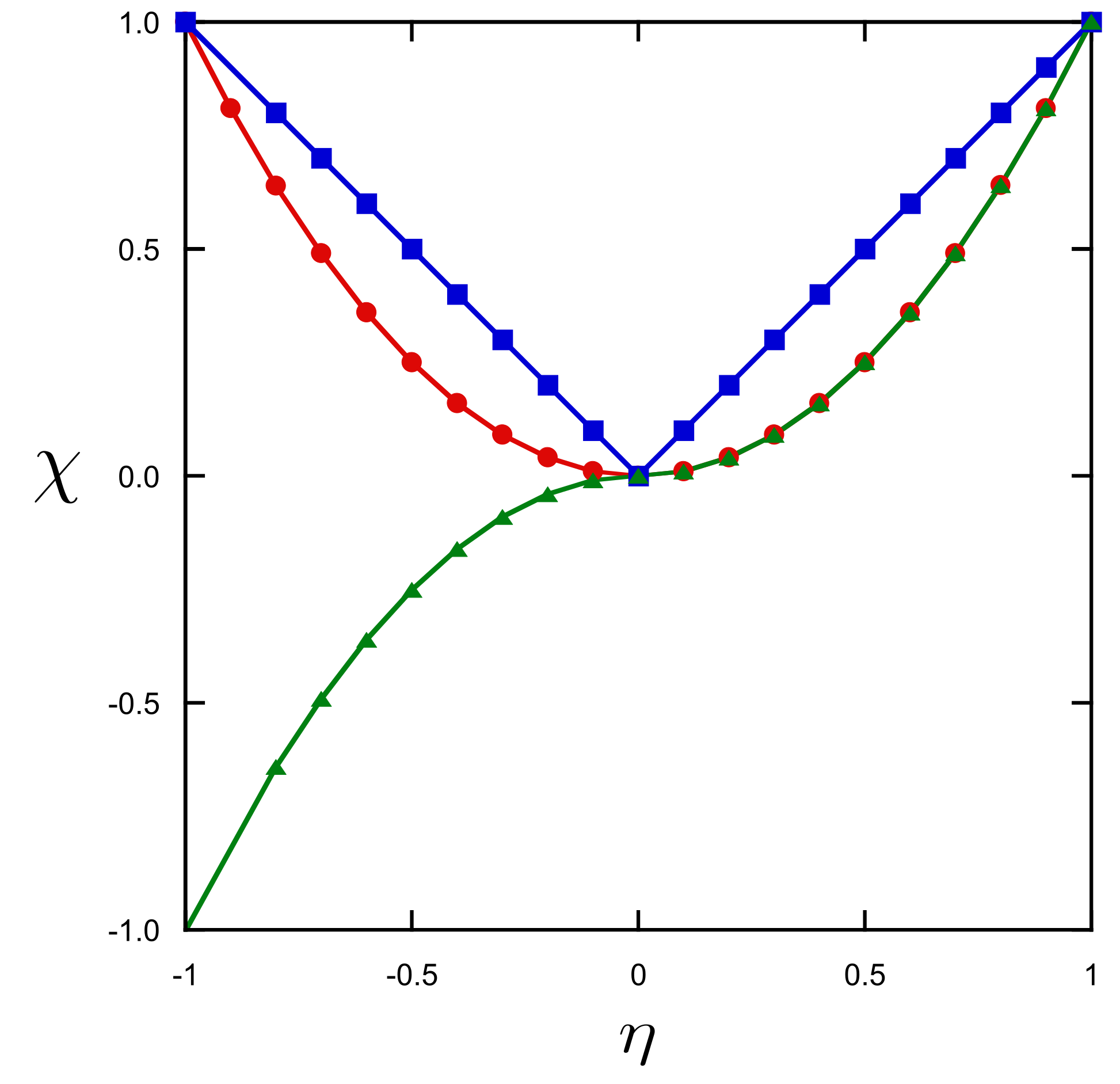}
      \caption{Comparison of the evolution of the different chiral measures as a function of the amplitude of the chiral distortion $\eta$ in Na$_3$AuO$_2$. Positive (negative) values of $\eta$ correspond to the condensation of the modes towards the $P4_12_12$ ($P4_32_12$) phase. Red dots, blue squares, and green triangles correspond to CCM, Hausdorff, and Helicity measures. The values of the different measures have been normalized to display a value of $1$ at the optimal amplitude of the chiral distortion ($\eta=1$).}
      \label{fig:comparison_chiral_measures} 
  \end{figure}
\end{center}

To compare those different measures, we plot in Fig.~\ref{fig:comparison_chiral_measures} the evolution of the CCM, Hausdorff, and helicity measures as a function of the distortion amplitude when going from the high symmetry $P4_2/mnm$ phase to both the $P4_12_12$ and $P4_32_12$ phases.
Those two phases are degenerate in energy and correspond to the two energy minima of the left and right-handed enantiomers that can be formed from the $P4_2/mnm$ phase.
The difference in the distortion pattern of those two enantiomeric phases is simply a change of sign, i.e., a change between right-handed and left-handed helical distortion.
As expected from the different definitions, the module of the Hausdorff distance shows a linear dependence with $\eta$. In contrast, the module of the CCM and the helicity measurements have a quadratic behavior, as already reported in Ref.~\cite{Felser-22}. 
Moreover, as anticipated, the helicity sign is reversed for the different enantiomers, whereas the CCM and Hausdorff distances are not.
Note that the CCM and Hausdorff distances could be multiplied by a sign representing the sense of rotation of the screw axis to differentiate between enantiomers. However, this measure will not serve to differentiate the degree of chirality of two different compounds due to the reference choice problem highlighted in Sec.~\ref{Sec:prob_CCM}.

Furthermore, when assessing the same calculations between the $P4_2/mnm$ phase and the $Cmcm$ intermediate achiral phase, which is found to be an intermediate achiral structure before the enantiomorphic groups~\cite{Fava-24}, we obtain a zero value from the helicity calculation.
Therefore, the helicity calculation appears robust in determining the chirality of helical structures and can be applied straightforwardly. Moreover, as one can see in Tab.~\ref{tab:data_gen}, when we apply this procedure to the case of the K$_3$NiO$_2$ and compare the results with the ones obtained for the Na$_3$AuO$_2$, we observe that the ratio between of the CCM's and the ratio of the helicities is constant suggesting the same helical structure for both compounds.
\section{The case of C\MakeLowercase{s}C\MakeLowercase{u}C\MakeLowercase{l}\textsubscript{3}}
\label{sec:CsCuCl3}

\begin{center}
  \begin{figure}[h]
     \centering
      \includegraphics[width=4cm]{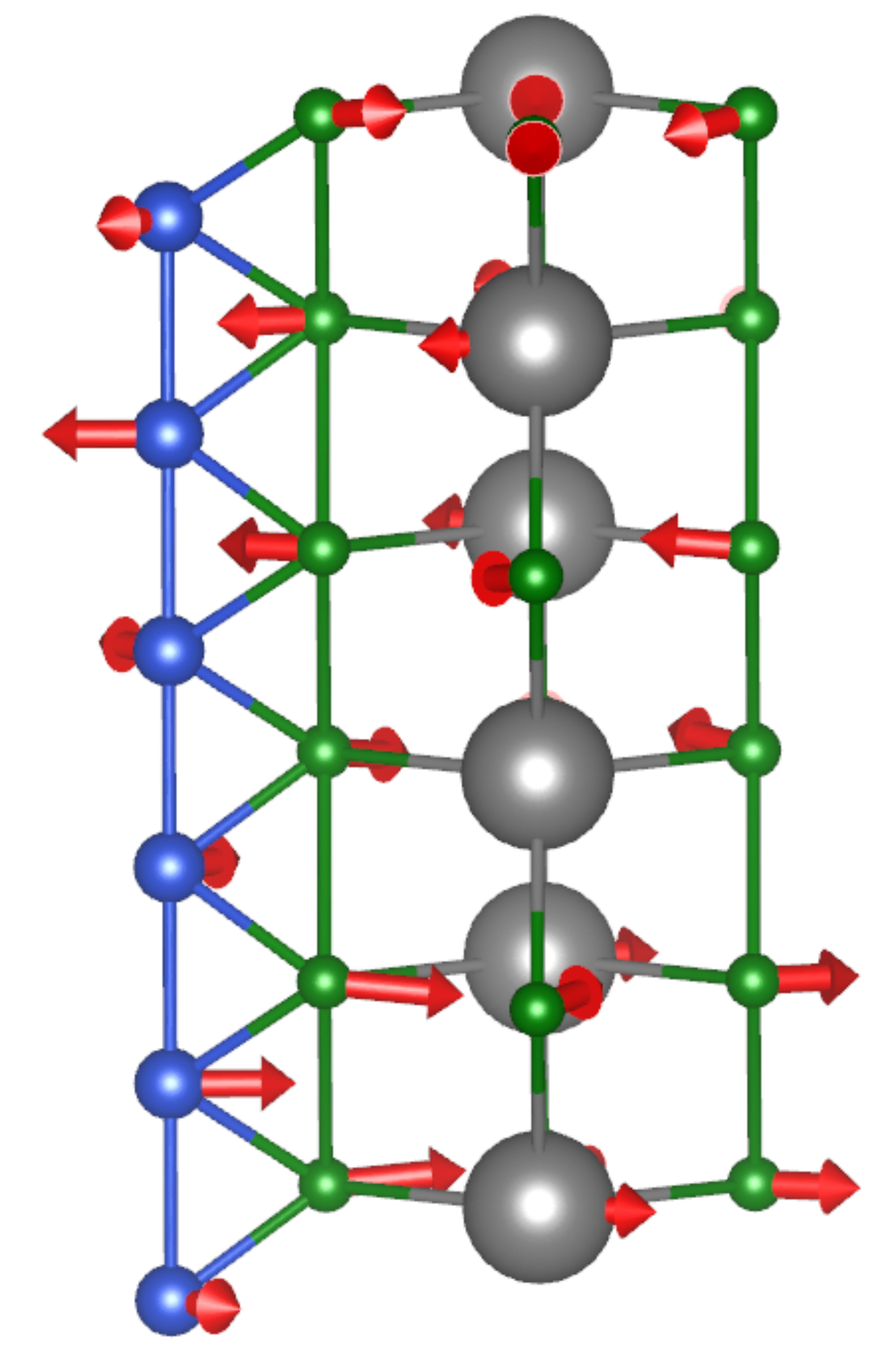}
      \caption{Schematic representation of the CsCuCl$_3$ where atoms occupy the high symmetry positions. Arrows indicate the direction of the displacements into the $P6_122$ chiral phase. Grey, blue, and green balls represent Cs, Cu, and Cl atoms.} 
      \label{fig:dist_CsCuCl3} 
  \end{figure}
\end{center}

Next, we will discuss the chiral transition in CsCuCl$_3$. This material undergoes a transition from a high-temperature, high-symmetry $P6_3/mmc$ phase to one of the enantiomorphic groups $P6_122$ or $P6_522$\cite{Koiso-96, Plakhty-09}. This transition is driven by a cooperative Jahn-Teller distortion as reported in Ref.~\cite{Kroese-74, Hirotsu-77}.
Figure \ref{fig:dist_CsCuCl3} illustrates a schematic representation of the CsCuCl$_3$ crystal, similar to our previous example. It shows the chiral crystal distortions, depicted as differences in atomic positions between the high-symmetry phase and one of the chiral phases. The displacement vector field demonstrates the right-handed helical structure. 
In Tab.~\ref{tab:data_gen}, the values of the different chirality measures can be encountered compared to other compounds.
As we can see from the data, if we compare the values of the different measures for the Na$_3$AuO$_2$ and the CsCuCl$_3$ cases, we can see that the ratio between the CCM is half of the ratio between the helicities, indicating that the CsCuCl$_3$ adopts a more pronounced helical structure at equal distortion.

\begin{table*}[hbtp]
\caption{\label{tab:data_gen}Values of the different chirality measures for the distinct compounds discussed in the work. Fractional units and normalization by the number of atoms in the unit cell have been performed to compute the CCM and helicity.}
\begin{tabular}{c  c  c  c  c}
\hline
   & Na$_3$AuO$_2$ &K$_3$NiO$_2$ & CsCuCl$_3$ & MgTi$_2$O$_4$ \\ \hline
CCM & 3.30e-3& 1.24e-3&5.66e-3&1.03e-4 \\
Hausdorff  &1.18e-1&7.35e-2 & 1.07e-1 &1.19e-2 \\
Helicity &7.75e-3 & 2.75e-3 &2.08e-2 &3.08e-4\\ 
\hline
\end{tabular}
\end{table*}
\section{The case of M\MakeLowercase{g}T\MakeLowercase{i}\textsubscript{2}O\textsubscript{4}}
\label{sec:MgTi2O4}
\begin{center}
  \begin{figure}[hbtp]
     \centering
      \includegraphics[width=7cm]{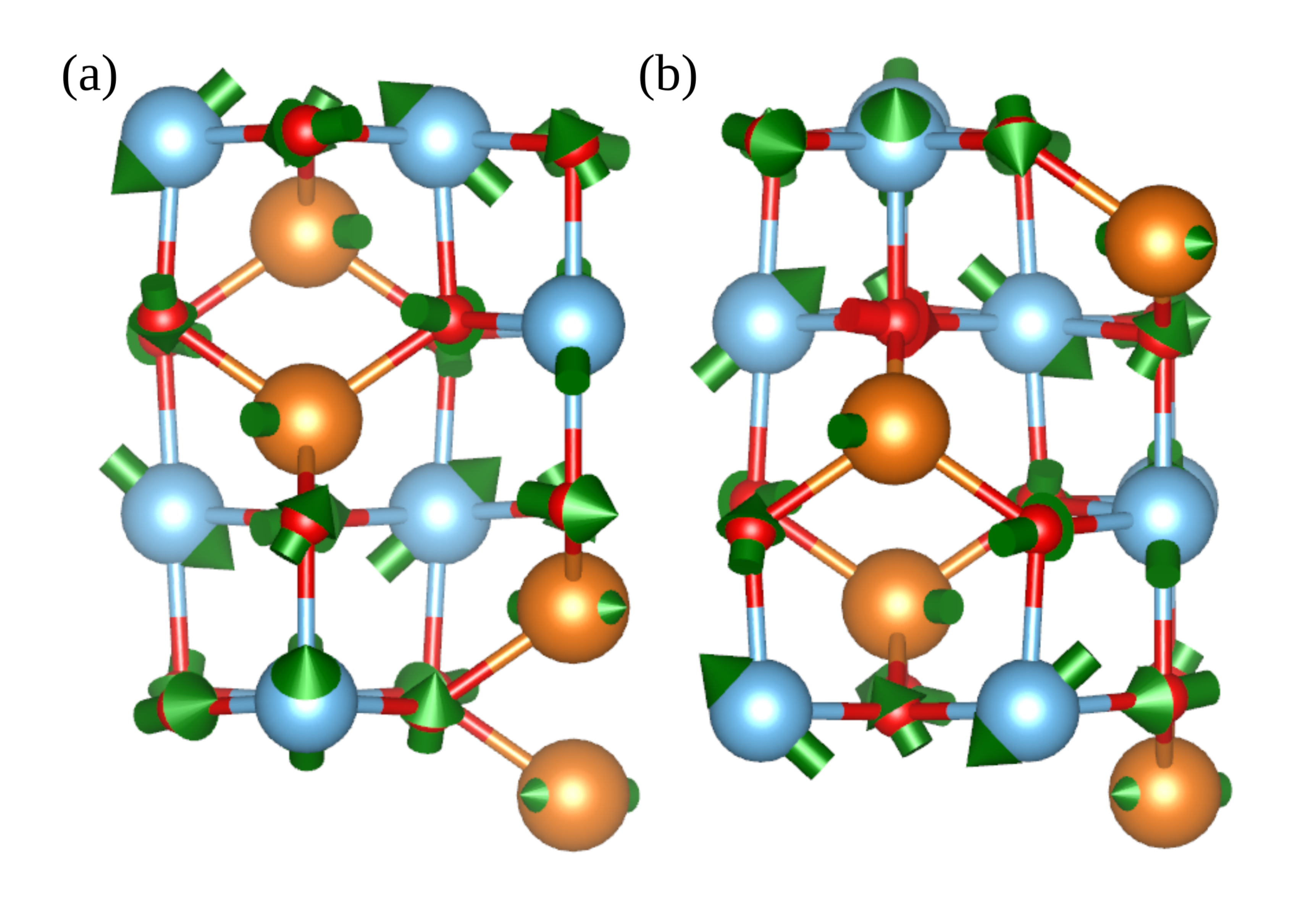}
      \caption{Schematic representation of the MgTi$_2$O$_4$. Atoms occupy the high symmetry undistorted positions in a (a) $P4_12_12$ or (b) $P4_32_12$ representation. Arrows indicate the direction of the displacements into their respective chiral phases. Orange, blue and red balls represent Mg, Ti and O atoms respectively.}
      \label{fig:dist_MgTi2O4} 
  \end{figure}
\end{center}

A crucial prerequisite for computing the helicity of a given structure is the ability to establish a one-to-one mapping between atomic positions in the high symmetry phase and those in the low symmetry phase, thereby defining the displacement vector field.
In previous examples, establishing this mapping has been straightforward. 
However, in the case of MgTi$_2$O$_4$, this is not so. 
The high-temperature phase exhibits a structure of $Fd-3m$ symmetry with 56 atoms in the conventional unit cell. In contrast, the chiral phase exhibits a $P4_12_12$ or $P4_32_12$ space group symmetry with 28 atoms in the unit cell~\cite{Schmidt-04,Ivanov-11}.
To establish such a mapping, we used the {\sc{ISODISTORT}} software\cite{Isodistort,Campbell-06}. 
Beginning with the high symmetry phase and setting the distortions that bring the system to the $P4_12_12$ or $P4_32_12$ phase to zero, we could derive a set of undistorted atomic positions that adhere to the desired symmetry.
Afterward, we provided to {\sc{AMPLIMODES}}\cite{Orobengoa-09,Perez-Mato-10} the undistorted and distorted atomic coordinates to obtain the set of displacements. 
In Fig.~\ref{fig:dist_MgTi2O4}, we present the atomic displacements obtained using this method. 

Similar to previous examples, the CCM and Hausdorff distances are equivalent for both enantiomers, while the helicity exhibits the same magnitude but opposite signs between them. 
The numerical values of the different chirality quantification methods can be found in Tab.~\ref{tab:data_gen}.
In comparison to the Na$_3$AuO$_2$ case, the ratio between the CCM values roughly equals the ratio between the helicities, suggesting that the two compounds present approximately the same chiral strength. 
In contrast to the previously discussed cases where a clear helical structure was observable along the $c$-direction, this compound exhibits a more evenly distributed helicity due to the lattice's transition from a face-centered cubic to a tetragonal structure. 
The corresponding distortion results in the absence of a preferred direction, making it more difficult to identify the helical structure of the compound visually. 
\section{Conclusions}
\label{sec:conclusions}
In this work, we have delved into various continuous chirality measures proposed in the literature. Utilizing the simple model system of a unit perovskite cell, we have examined the challenges these measures pose when applied to extended solids. 
For the case of the CCM or Hausdorff distances, in addition to the difficulties associated with employing a position-based operator in periodic systems~\cite{Resta-98}\footnote{The position operator does not commute with the Hamiltonian.}, not discussed in the text, we have identified the critical issue of selecting an appropriate reference, which can often be non-trivial.
Even in the case of small chiral distortions, relying on the high-symmetry phase as an accurate reference may not suffice, as the deformation could involve multiple non-chiral modes that only become chiral when coupled together as exemplified by the rotation (axial-like) and off-centering (vector) distortions.
Consequently, much important information about the distortion can remain implicitly hidden even with the right reference selection.
Moreover, because these methods rely on computing distances, which are positive definite quantities, they cannot discriminate between the different enantiomorphic structures of a given compound.
An interesting venue to explore would be to create tuples of chiral measures and increasing the number of degrees of freedom in order to have a better description of the distortion. However, there are strong mathematical evidence to believe that it will be insufficient to solve the chiral contentedness and consequent false zeros problem~\cite{Weinberg-97,Vavilin-22}.

Besides the analysis of the various chirality measures in the literature~\cite{petijean2003}, we have proposed using a novel pseudoscalar function, the helicity, to quantify the handedness of solids that undergo a continuous transition from an achiral to a chiral enantiomorphic space group. 
Borrowing the definition from hydrodinamics~\cite{Moffat-92,Moffatt2014}, and its extension to discrete fields in periodic solids~\cite{Shafer-18,Junquera-23}, this quantity can be directly computed from the eigenvector that brings the system from the high-symmetry non-chiral phase to the low-symmetry chiral one.
We have demonstrated the method's robustness in measuring the handedness of a given distortion in enantiomorphic groups that do not suffer the false zeros problem. 
In such groups, helicity not only yields zero values for non-chiral transformations but also provides finite values for chiral structures, with equal magnitude and opposite signs for each enantiomorph. 
Such approach has unveiled the handedness of different compounds like Na$_3$AuO$_2$, CsCuCl$_3$ or MgTi$_2$O$_4$. The use of helicity as defined in the article could therefore be incorporated as a potential descriptor for the high-throughput characterization of materials, similar to how the CCM has been utilized for organic molecules~\cite{Zahrt2019} as well as interesting to characterize chiral phonons~\cite{Ueda-23}.
While effective in measuring the handedness of crystals with enantiomorphic space groups (the 11 enantiomorphic pairs), it is important to note that chiral crystals that are within the 43 non-enantiomorphic space groups present chiral-connectedness such that this approach may be inadequate in those cases enhancing the well-known difference between chirality and handedness~\cite{Felser-22}. We note that the method is limited to measure the handedness of systems that exhibit a displacive transition from an achiral reference to a chiral state. In order to extend this mechanism to a broader range of compounds, a minimal supergroup search can be conducted, analogous to the approach used for identifying displacive ferroelectrics~\cite{Kroumova-00}. A comprehensive study detailing the workflow and introducing new candidates for soft phonon mode-driven displacive chiral phase transitions is underway~\cite{Gomez-25}.
Additionally, Appendix~\ref{app:Fourier_hel} presents a perspective on a potential reciprocal-space formulation of the helicity operator. Although not implemented in this work, this approach could inspire further advancements in the study of chiral distortions, potentially enabling its integration into density functional perturbation theory and finite-wavelength techniques~\cite{royo2019}.

Even though the helicity described in this article remains reliant on a position-based operator and thus susceptible to common challenges in dealing with periodic systems~\cite{Resta-98}, we hope that our proposed approach will facilitate a more systematic quantification of crystal handedness of enantiomorphic space groups. We believe that the helicity measurement described here shares some similarities to the polarization quantification problem in ferroelectrics. The most evident example is that we aim to measure to what extent a chiral crystal deviates from its achiral high symmetry phase. Besides, some of the issues discussed in the manuscript, as the reference problem, also existed (and were solved by the modern theory of polarization~\cite{King-Smith}) in the case of the polarization and periodic boundary conditions. We hope that the present manuscript will stimulate further work towards refining the quantification of structural chirality.
For instance, it would be interesting to establish a relation between helical structures' helicity from enantiomorphic groups and their optical activity, e.g., from density functional theory as the optical activity calculation has been recently implemented~\cite{zabalo2023}.
\acknowledgments
F.G.-O., M.F. and E.B. acknowledge the Fonds de la Recherche Scientifique (FNRS) for financial support, the PDR project CHRYSALID No.40003544 and the Consortium des Équipements de Calcul Intensif (CÉCI), funded by the F.R.S.-FNRS under Grant No. 2.5020.11 and the Tier-1 Lucia supercomputer of the Walloon Region, infrastructure funded by the Walloon Region under the grant agreement No. 1910247. F.G.-O. and E. B. also acknowledge support by the European Union’s Horizon 2020 research and innovation program under Grant Agreement No. 964931 (TSAR). F.G.O. also acknowledges financial support from MSCA-PF 101148906 funded by the European Union and the Fonds de la Recherche Scientifique (FNRS) through the grant FNRS-CR 1.B.227.25F
The work at West Virginia University was supported by the U.S. Department of Energy (DOE), Office of Science, Basic Energy Sciences (BES), 
under Award DE‐SC0021375. This work used Bridges2 and Expanse at the Pittsburgh Supercomputer and the San Diego Supercomputer Center through allocation DMR140031 from the Advanced Cyberinfrastructure Coordination Ecosystem: Services \& Support (ACCESS) program, which National Science Foundation supports grants 2138259, 2138286, 2138307, 2137603, and 2138296.
\appendix
\section{Reciprocal space formulation of the Helicity}
\label{app:Fourier_hel}
This brief section explores a potential alternative formulation of the helicity operator in reciprocal space. 

Notably, the value of the Fourier transform of the helicity density operator at the $\Gamma$ point directly corresponds to the helicity
\begin{align}
    \hat{H}(\vec{k}=0)&=\int \vec{v}\cdot(\nabla\times\vec{v})\cdot e^{i\vec{0}\vec{r}}d^3\vec{r}\nonumber\\
    &=\int \vec{v}\cdot(\nabla\times\vec{v})\cdot 1d^3\vec{r}=\mathcal{H}.\nonumber
\end{align}
Now, we only have to compute the Fourier transform of the helicity density operator $\hat{H}(\vec{r})=\vec{v}\cdot(\nabla\times\vec{v})$. Applying the convolution theorem~\cite{vanDijk-13}, the Fourier transform of the product of two vectorial functions corresponds to the component-wise convolution of their respective Fourier transforms. Therefore, the following equality holds
\begin{equation}
\mathcal{F}\left[\vec{v}\cdot(\nabla\times\vec{v})\right]=\mathcal{F}\left[\vec{v}\right]\dot{\ast}\mathcal{F}\left[\nabla\times\vec{v}\right].
\end{equation}
Moreover, the Fourier transform of the curl of a function can be rewritten in the following way~\cite{vanDijk-13}
\begin{equation}
    \mathcal{F}[\nabla\times\vec{v}]=i\vec{k}\times\mathcal{F}\left[\vec{v}\right],
\end{equation}
which offers a concise expression for the reciprocal-space formulation of the operator discussed in the main text.
\end{document}